\documentclass[aps,preprint,prl]{revtex4}
\usepackage{graphicx,amssymb,amsmath}
\usepackage{graphicx}% Include figure files
\usepackage{dcolumn}% Align table columns on decimal point
\usepackage{bm}% bold math

\newcommand{\nt}{$\nu_T=1$}
\newcommand{\dl}{$d/\ell$}

\newcommand{\bpar}{$B_{||}$}

\newcommand{\Sxx}{$\sigma_{xx}^{||}$}
\newcommand{\SxxCF}{$\sigma_{xx}^{CF}$}

\newcommand{\dsas}{$\Delta_{SAS}$}
\begin{document}

\title{Exciton Condensation in Bilayer Quantum Hall Systems}

\author{J.P. Eisenstein}

\affiliation{Condensed Matter Physics, California Institute of Technology, Pasadena, CA 91125}

\date{\today}

\begin{abstract} The condensation of excitons, bound electron-hole pairs in a solid, into a coherent collective electronic state was predicted over 50 years ago.  Perhaps surprisingly, the phenomenon was first observed in a system consisting of two closely-spaced parallel two-dimensional electron gases in a semiconductor double quantum well.  At an appropriate high magnetic field and low temperature, the bilayer electron system condenses into a state resembling a superconductor, only with the Cooper pairs replaced by excitons comprised of electrons in one layer bound to holes in the other. In spite of being charge neutral, the transport of excitons within the condensate gives rise to several spectacular electrical effects. This article describes these phenomena and examines how they inform our understanding of this unique phase of quantum electronic matter.

\end{abstract}

\maketitle
\section{Introduction and Scope}
It is by now well established that when two parallel two-dimensional electron systems are brought close together in the presence of a large magnetic field, electron-electron interactions both within and between layers can conspire to drive a transition into a remarkable new phase of quantum electronic matter.  In this new phase electrons lose their memory of which layer they are in, and instead inhabit both layers equally.  It is most remarkable that this ``which-layer uncertainty" does not hinge on the presence of a finite interlayer tunneling amplitude but instead develops spontaneously as the system reaches for the correlated state with the lowest Coulomb interaction energy.

To date, this new phase has been experimentally observed only when the total electron density in the bilayer, $n_{T}$, matches the degeneracy $eB/h$ of the lowest single spin-resolved Landau level created by the magnetic field, $B$.  This situation therefore corresponds to total Landau level filling factor \nt.  Interestingly, it is only the $total$ electron density which must match the level degeneracy; the apportionment of $n_T=n_1+n_2$ between the two layers is not important (to a first approximation).  Similar phases are expected to exist at $\nu_T =3$, 1/3, and elsewhere, but these have so far eluded detection.

There are multiple equivalent languages in which the physics of the new phase can be described.  Notable among these are the pseudospin ferromagnetism language and the exciton condensation language.  In the pseudospin ferromagnetism picture, electrons definitely in one layer are declared pseudospin ``up", while those definitely in the other layer are pseudospin ``down".  In the condensed phase of a balanced bilayer ($n_1=n_2=n_T/2$) at small layer separation, exchange interactions favor a state in which the pseudospins of all electrons are parallel and lie in the $x-y$ plane of pseudospin space (the electrons thus occupying both layers simultaneously).  The system resembles a metallic easy-plane ferromagnet.  Alternatively, in the exciton condensation language one takes advantage of the finite number of states within a Landau level to focus on the empty states rather than the filled ones.  The condensed phase may then be viewed as electrons in the lowest Landau level of one layer bound to holes in the lowest Landau level of the other layer.  As long as \nt, the number of holes in the one layer always matches the number of electrons in the other.  In close analogy to the Cooper pairs in a standard BCS superconductor, the number of these electron-hole pairs, or excitons, is not a good quantum number.  It is the attractive interlayer interaction between the electrons and holes that allows the bilayer system to condense into an especially low energy state, one not available to the individual layers themselves.

The pseudospin ferromagnetism and exciton condensation pictures are completely equivalent and can be used interchangeably.  For the purposes of the present article however, we will predominantly employ the exciton picture.  The reason for this is simple, if largely aesthetic:  The most direct experimental demonstrations of the unusual physics of the \nt\ condensed phase involve driving electrical currents in opposite directions through the two two-dimensional electron gas layers.  Such counterflowing currents are readily envisioned as uniform flows of charge neutral excitons.  Signature experimental results, such as the vanishing of the Hall resistance in counterflow, are rendered intuitive in the exciton language.

The condensed \nt\ bilayer system at small layer separation exhibits an energy gap to its charged excitations and displays the quantized Hall effect (QHE).  This contrasts sharply with the situation at larger layer separation where (in the balanced case) no energy gap nor Hall plateau is observed.  Remarkably, this transport phenomenon, which is observed when parallel currents flow in the two layers, is among the $less$ interesting aspects of the excitonic phase.

The most interesting, and so far unique, aspects of the condensed \nt\ phase stem from the presence of a condensate degree of freedom.  In essence, the contrast between an ordinary quantum Hall system and the present \nt\ bilayer system is analogous to the distinction between a semiconductor and a superconductor.  Both the semiconductor and the superconductor display an energy gap and current-carrying excited states, but only the superconductor possesses a Cooper pair condensate capable of transporting current without dissipation.  Similarly, both ordinary QHE states and the present bilayer \nt\ state display a gap to charged excitations, but only the \nt\ system possesses an exciton condensate capable of (neutral) superfluid transport.  Indeed, it is the presence of the condensate that has motivated almost two decades of theoretical and experimental study of the bilayer \nt\ system.  It is also the main focus of the present article.

The scope of this paper is limited to experimental investigations of the \nt\ bilayer exciton condensate by means of electrical transport.  As such, no attempt to review the many very interesting experimental studies of the transition into the excitonic phase from the weakly-coupled state at large layer separation has been made, nor have the numerous important results of alternative experimental probes (e.g. optical, thermoelectric, etc.) been included.  A comprehensive review of the large associated theoretical literature on the \nt\ problem has wisely been left for an expert.

\section{Snapshots of Exotic Phenomena at $\nu=1$}

\begin{figure}
\begin{center}
\includegraphics[width=6in] {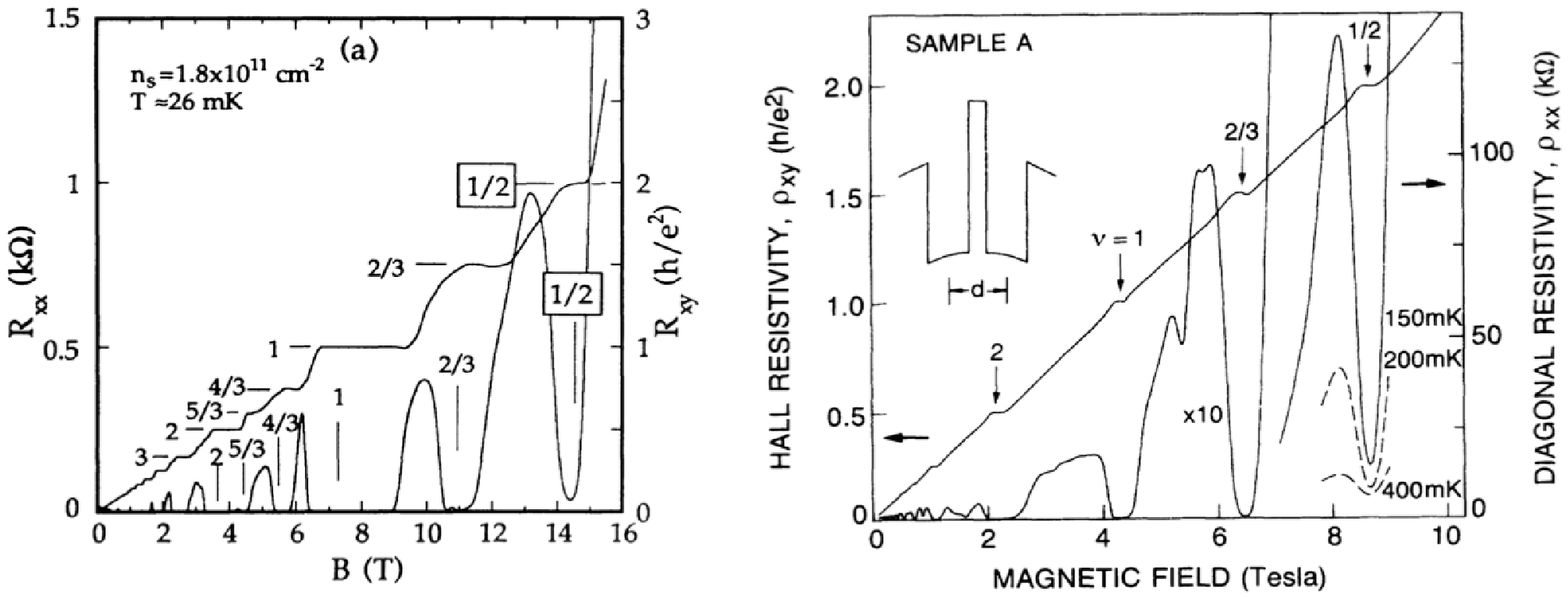}
\end{center}
\caption{Discovery of quantized Hall states at \nt\ and $\nu_T = 1/2$ in bilayer electron systems.  In the left panel the bilayer system is realized in a wide single quantum well, while on the right the electrons reside in a double quantum well.  After Suen {\it et al.} \cite{suen92} and Eisenstein {\it et al.} \cite{eisenstein92}.}
\label{fig1}
\end{figure}

The existence of a quantized Hall plateau and an associated vanishing longitudinal resistivity at \nt\ (see Fig. \ref{fig1} for early examples \cite{suen92,eisenstein92}) demonstrates that closely-spaced bilayer 2D electron systems at this filling factor possess an energy gap to charged excitations.  In this the system is no different than any other quantized Hall system.  However, the existence of a condensate capable of coherent transport wholly independent of the charged excitations is not revealed by the data shown in Fig. \ref{fig1}.  The condensate degree of freedom is hidden from conventional electrical measurements based upon parallel current flow in the two layers.  To couple to the condensate, experiments involving anti-parallel, or counterflowing, currents must be performed.  Two such experiments are discussed here in thumbnail form.  Experiments such as these require separate electrical connections to the two layers in the bilayer sample.  Fortunately, a robust technique for establishing such contacts was developed in 1990 \cite{eisenstein90}.

\subsection{Josephson-like Tunneling Anomaly}
\begin{figure}
\begin{center}
\includegraphics[width=6in] {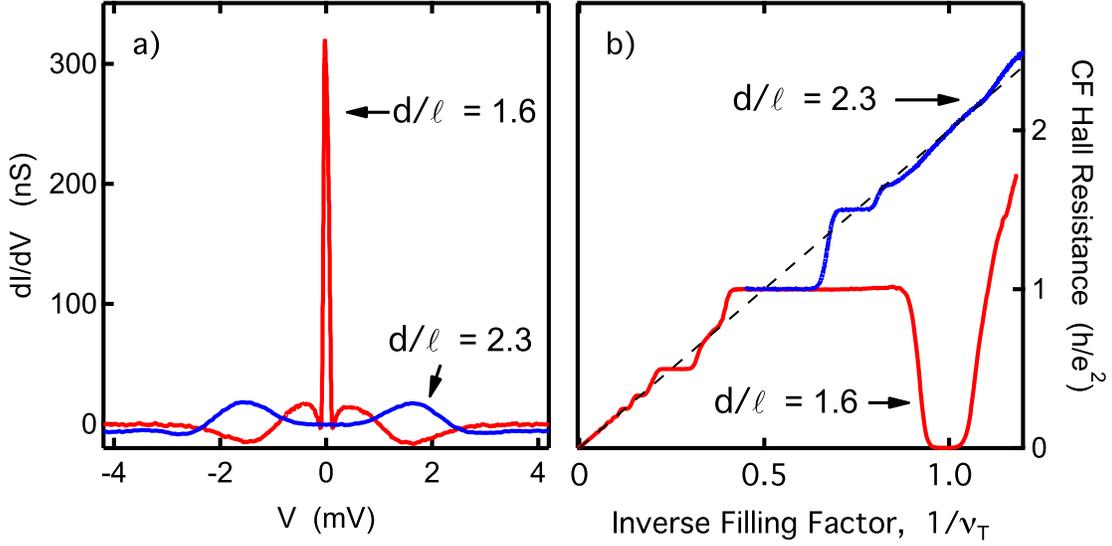}
\end{center}
\caption{Signatures of exciton condensation at \nt\ in a bilayer 2D electron system at low temperatures ($T\leq 50$ mK).  a) Interlayer tunneling conductance $dI/dV$ versus interlayer voltage $V$ at \nt.  At $d/\ell=2.3$ (blue trace) $dI/dV$ is heavily suppressed around zero bias.  At $d/\ell=1.6$ (red trace) exciton condensation has generated a huge, highly resonant peak in the tunneling conductance.  b) Hall resistance in counterflow versus inverse filling factor $1/\nu_T$ for a single sample but at different total electron densities $n_T$.  At high density (blue trace), where $d/\ell=2.3$ at \nt, the CF Hall resistance essentially follows the classical Hall line (dashed) in the vicinity of \nt.  However, for low density (red trace), where $d/\ell=1.6$ at \nt, exciton condensation has quenched the CF Hall resistance around \nt\ even as it remains essentially unaffected at other filling factors.  After Spielman {\it et al.} \cite{spielman00} and Kellogg {\it et al.} \cite{kellogg04}.}
\label{thumbs}
\end{figure}

Figure \ref{thumbs}a shows two interlayer tunneling conductance traces obtained from a single balanced bilayer 2DES sample.  Each trace was recorded at \nt,  albeit at different total densities $n_T$ and magnetic fields.  (Adjusting the density is enabled by electrostatic gates deposited on the sample's top and backside surfaces.)  For the red and blue traces shown the effective interlayer separation \dl, where $d$ is the center-to-center distance between the two GaAs quantum wells (here $d=28$ nm), and $\ell=(\hbar/eB)^{1/2}$ is the magnetic length, is \dl$~=1.6$ and 2.3, respectively.  For the high density, \dl$~=2.3$ data, the tunneling conductance near zero interlayer voltage is very small.  Indeed, the zero bias tunneling conductance is heavily suppressed by the strong Coulomb correlations characteristic of Landau quantized {\it single layer} 2D electron systems.  This suppression reflects a Coulomb pseudogap in the tunneling density of states which denies rapidly injected tunneling electrons access to any low energy, highly correlated states the 2DES may possess.  In contrast, the red trace in Fig. \ref{thumbs} shows a strong, highly resonant peak in the tunneling conductance at zero bias.  This peak, which appears to be confined to the same range of parameters (\dl, temperature, magnetic field, etc.) as the \nt\ QHE, is fundamentally rooted in the ``which layer'' uncertainty characteristic of the excitonic phase.  The strong interlayer correlations built into the phase ensure that there is no energy cost associated with the transfer of an electron from one layer to the other.  Crudely speaking, an electron about to tunnel from one layer is assured of the existence of a hole immediately opposite it in the other layer.  The tunneling conductance peak may be viewed as an indirect signature of counterflow, or excitonic, superfluidity.  An electron tunneling between layers momentarily creates a localized charge build-up in one layer and a corresponding deficit in the other.  For widely separated layers these twin defects relax independently and only very slowly owing to the small conductivity $\sigma_{xx}$ of 2D electrons in large magnetic fields.  However, once the excitonic phase is established at smaller layer separation, the very high counterflow conductivity of its condensate is the perfect vehicle for rapidly relaxing this antisymmetric bilayer charge defect.  

The sharply resonant peak in the tunneling conductance at \nt\ signals a nearly discontinuous jump in the tunneling current at zero interlayer voltage. This is obviously reminiscent of the dc Josephson effect in superconducting tunnel junctions.  How close this connection really is remains a subject of intense experimental and theoretical interest.  

\subsection{Vanishing Hall Resistance in Counterflow}
Figure \ref{thumbs}b shows another dramatic consequence of exciton condensation at \nt.  The Hall resistance measured in one of the two 2D layers is plotted vs. inverse total filling factor $1/\nu_T$.  As in Fig. \ref{thumbs}a, the two curves derive from different total electron densities $n_T$ and therefore different effective layer separations \dl\ at \nt.  Unlike an ordinary Hall effect measurement in which equal, co-directed currents flow in the two layers, in this case the individual layer currents are equal in magnitude, but opposite in direction.  This is, therefore, a counterflow (CF) experiment.  As the magnetic field (and hence $1/\nu_T$) is increased from zero, a series of Hall plateaus is observed.  These plateaus correspond to the familiar quantized Hall states present in each layer separately.  The fact that the current is being returned via the opposite layer is of no consequence.  However, around \nt\ the CF Hall resistance behaves very differently in the two curves.  For the high density data (blue), where $d/\ell = 2.3$ at \nt, the Hall resistance remains close to the classical Hall line.  In contrast, for the low density data, where $d/\ell = 1.6$ at \nt, the CF Hall resistance drops essentially to zero at \nt.  (Importantly, the Hall voltage measured in the other layer also falls close to zero around \nt.)  As with the tunneling anomaly discussed above, this vanishing of the Hall resistance in counterflow only occurs when the \nt\ excitonic QHE state is present.  A facile explanation of this remarkable result is evident: Counterflowing currents in the two layers can be transported by the condensate which consists entirely of charge neutral excitons.  Being neutral the excitons experience no Lorentz force and thus the counterflow Hall voltage vanishes.  As will become clear in subsequent sections, this simple explanation is convincingly supported by the most recent experiments.

\section{Bilayer Quantum Hall Effects}
In a seminal 1983 paper, Halperin \cite{halperin83} extended the argument given by Laughlin \cite{laughlin83} for the fractional quantized Hall effect (FQHE) to two-dimensional electron systems (2DESs) possessing a discrete internal quantum number.  While the most obvious such quantum number is the electron spin (and Halperin's wavefunctions do accurately approximate certain spin unpolarized FQHE states), layer index in bilayer 2DES systems is another possibility.  Subsequent work by Haldane and Rezayi and by Pietilainen and Chakraborty made the case for new, intrinsically bilayer, FQHE states at $\nu_T = 1/2$ and \nt, respectively \cite{haldane87,chakraborty87}.

On the experimental side, the fabrication of closely-spaced double layer 2D electron systems of high quality presented challenges beyond those already overcome in the growth of high mobility single layer systems.  Nevertheless, by around 1990 GaAs-based bilayer 2DESs with mobilities around $10^6$ cm$^2$/Vs had been obtained.  Two main architectures were pursued: Wide single quantum wells in which electrostatic effects produce a ground subband wavefunction with a pronounced dumbbell shape, and true double quantum wells consisting of two thin GaAs layers embedded in the alloy Al$_x$Ga$_{1-x}$As.  Since it is generally desirable to keep single particle tunneling small in comparison with interlayer Coulomb interactions, the latter geometry has dominated experiments on the \nt\ state, both in electron-electron and hole-hole bilayers.  The need for weak tunneling and small layer separation obviously conflict. The result has been structures in which the alloy barrier layer separating the two quantum wells has a high aluminum concentration, Al$_{0.9}$Ga$_{0.1}$As being typical, in order to increase the tunnel barrier height.  Unfortunately, the high reactivity of aluminum, which concentrates residual impurities in the barrier layer, contributes to the relatively low mobilities of contemporary double layer 2D electron systems used for \nt\ research.

The first experimental evidence for unusual quantized Hall states in bilayer electron systems was obtained in 1992 \cite{suen92,eisenstein92}.  Much of the original excitement was focused on the presence of a Hall plateau at $\rho_{xy}=2h/e^2$.  This plateau occurs at total Landau level filling $\nu_T=1/2$, which corresponds to individual 2D layer fillings of $\nu_1=\nu_2=1/4$.  This excitement was due, in part, to the obvious violation of the famous ``odd-denominator rule'' governing virtually all previously known FQHE states \cite{willett87}.  Like the famous $\nu = 1/3$ FQHE in single layer systems, the $\nu_T=1/2$ bilayer FQHE is due entirely to electron-electron interactions.

The same bilayer 2D systems that exhibited a $\nu_T=1/2$ quantized Hall plateau at $\rho_{xy}=2h/e^2$ also displayed one at \nt\ where $\rho_{xy}=h/e^2$.  For a balanced bilayer the individual filling factors are therefore $\nu_1=\nu_2=1/2$.  Owing to a remarkable broken symmetry, spontaneous interlayer phase coherence (the source of ``which layer'' uncertainty), the \nt\ state has turned out to be far more interesting than the $\nu_T=1/2$ state.  Indeed, the \nt\ state exhibits several remarkable transport properties not shared by any other experimentally observed quantum Hall state.  Figure  \ref{fig1} displays the early transport data which first revealed the \nt\ and $\nu_T=1/2$  bilayer QHE states \cite{suen92,eisenstein92}.

\section{Pseudoferromagnetism and Exciton Condensation at \nt}

In the pseudospin language \cite{pstheory,fertig89,wen92,yang94,moon95,smg_ahm97} the ground state wavefunction for the balanced \nt\ bilayer at small layer separation is well-approximated by

\begin{equation}
|\Psi\rangle = \prod_k \frac{1}{\sqrt2}(|\uparrow\rangle + e^{i \phi}|\downarrow\rangle)\otimes|k \rangle
\label {eq1}
\end{equation}
where $|\uparrow\rangle$ and $|\downarrow\rangle$ are pseudospin eigenstates representing electrons definitely in the ``top'' and ``bottom'' 2D layers, respectively, while the product runs over all the momentum eigenstates $|k\rangle$ in the lowest Landau level.  (The true spin of the electrons is not indicated in Eq. \ref{eq1} and is assumed to be fully polarized along the magnetic field direction.) It is clear from the structure of Eq. \ref{eq1} that $|\Psi\rangle$ represents a single fully-filled Landau level of electrons, each of which is in a coherent linear superposition of top and bottom layer states (and thus exhibiting ``which layer'' uncertainty).  The phase $\phi$ is the same for all electrons and, in the absence of extrinsic symmetry-breaking fields, is arbitrary.  The state $|\Psi\rangle$ is (pseudo-) ferromagnetically ordered with total moment lying in the $x-y$ plane, inclined by the angle $\phi$ relative to the $x$-axis.  The onset of ferromagnetic order (in this case a breaking of U(1) symmetry) is spontaneous, driven by strong intra- and interlayer exchange energies. Interlayer tunneling, which is never completely absent, explicitly breaks the U(1) symmetry by adding, in effect, a pseudo-magnetic field along the $x$ axis of pseudospin space.  Tunneling thus does not disrupt the ferromagnetic order, but it does tend to orient the moment by favoring symmetric ($\phi = 0$) bilayer eigenstates. 

In second-quantized notation the ground state given in Eq.\ref{eq1} is written
\begin{equation}
|\Psi\rangle = \prod_k \frac{1}{\sqrt2}(c^\dagger_{k,T} + e^{i \phi} c^\dagger_{k,B})|0 \rangle
\label {eq2}
\end{equation}
where $c^\dagger_{k,T}$ ($c^\dagger_{k,B}$) creates an electron of momentum $k$ in the lowest Landau level in the top (bottom) layer out of a vacuum state $|0\rangle$ containing no conduction band electrons in either layer.  Alternatively, if a vacuum state $|0'\rangle$ consisting of a filled lowest Landau level in the top layer but no electrons in the bottom layer is used, then $|\Psi\rangle$ becomes 
\begin{equation}
|\Psi\rangle = \prod_k \frac{1}{\sqrt2}(1 + e^{i \phi} c^\dagger_{k,B}c_{k,T})|0' \rangle.
\label {eq3}
\end{equation}
Written in this way the exciton condensation picture is clearly revealed.  The operator $c^\dagger_{k,B}c_{k,T}$ simultaneously creates a hole in the top layer and an electron in the bottom layer, each of momentum $k$ in the lowest Landau level.  As expected, the average number of such excitons in the (balanced) \nt\ state is one-half the total number of electrons in the bilayer.  The similarity of Eq. \ref{eq3} to the BCS ground state of a conventional superconductor is also obvious \cite{bcs57}.  In that case the vacuum is a filled Fermi sea; here it is filled lowest Landau level in one layer.  The usual BCS coherence factors $u_k$ and $v_k$ which encode the momentum dependence of the pairing are absent here; owing Landau level degeneracy, all $k$ states are equivalent and the coherence factors are simply $1/\sqrt{2}$.  Finally, of course, Cooper pairs carry charge 2$e$ while excitons are electrically neutral.

There are a variety of excitations of the coherent \nt\ state.  Vortices in the phase field $\phi$, known as merons and anti-merons, carry both real and topological charge.  These gapped excitations are responsible for the ordinary quantized Hall effect and charge transport in the bulk of the system.  At long wavelengths there are neutral excitations governed by the effective Hamiltonian
\begin{equation}
H = \int {d^2r \lbrace \frac{\rho_s}{2}(\nabla\phi)^2 + \beta m_z^2 -\frac{t}{2\pi\ell^2} \cos \phi \rbrace }.
\label{eq4}
\end{equation}
The first term reflects the exchange energy cost of spatial variations in $\phi$, while the second term expresses the capacitive energy penalty associated with out-of-plane excursions of the local pseudospin moment. (The $z$-component of the pseudospin, $m_z$, is just the difference in layer filling factors, $m_z=\nu_1-\nu_2$.)  The pseudospin stiffness (or superfluid density) $\rho_s$ has been estimated theoretically, and $\beta^{-1}$ is proportional to the capacitance between the layers \cite{moon95,smg_ahm97}.  The last term is the explicit U(1) symmetry breaking term due to ordinary tunneling, with $t$ the tunneling matrix element.  Among other things, Eq. \ref{eq4} leads to the linearly dispersing pseudospin waves \cite{spinwavegap} first predicted by Fertig \cite{fertig89} and observed by Spielman {\it et al.} \cite{spielman01}.

\section{Phase Transition at \nt}

\begin{figure}
\begin{center}
\includegraphics[width=5in] {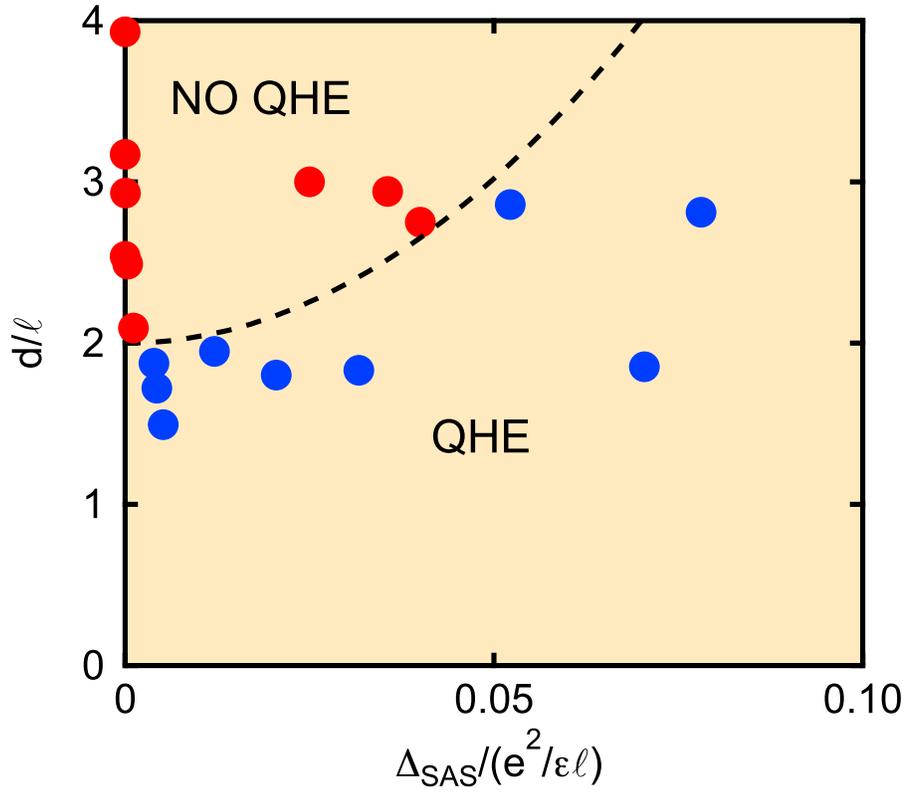}
\end{center}
\caption{Phase diagram of the \nt\ bilayer 2DES system in the \dl\ -- $\Delta_{SAS}/(e^2/\epsilon\ell)$ plane.  Blue circles: Bilayer samples which display a quantized Hall plateau at $\rho_{xy} = h/e^2$ at \nt.  Red circles: Samples which do not display a quantized Hall plateau at \nt.  This figure strongly suggests that the \nt\ QHE persists even in the limit of zero tunneling.  After Murphy {\it et al.} \cite{murphy94}.}
\label{phasediag}
\end{figure}

Figure \ref{phasediag} shows the first attempt to establish a phase diagram for the \nt\ bilayer system \cite{murphy94}.  Each red or blue circle in the figure gives the coordinates of a distinct bilayer sample in a dimensionless layer separation -- tunneling strength plane.  The center-to-center quantum well separation $d$ is normalized by the magnetic length $\ell$ at \nt, while the tunneling strength is parameterized by the splitting \dsas\ ($= 2t$) between the lowest symmetric and anti-symmetric eigenstates of the double well structure, normalized by the mean Coulomb energy $e^2/\epsilon\ell$ at \nt.  The blue circles indicate samples which do display a quantized Hall effect at \nt\ (with $\rho_{xy} = h/e^2$), while the red circles denote samples that do not.  The figure reveals a clear separation between a QHE regime at small \dl\  and a non-QHE regime at large \dl.   The figure also shows that the domain occupied by the QHE phase grows smoothly as the tunneling strength is increased.  This makes sense given the discussion in the previous section; tunneling strengthens the interlayer phase coherence which otherwise develops spontaneously at small \dl.  (Obviously, for sufficiently large \dsas\ the system is effectively a single 2D layer, with \nt\ corresponding to one filled Landau level of spin-polarized symmetric state electrons.  In this regime, the single particle tunnel splitting \dsas\ is alone sufficient to establish a QHE \cite{spin1}).  Most importantly, Fig. \ref{phasediag} strongly suggests that the bilayer \nt\ QHE persists even in the limit of zero tunneling.  In this limit, there appears to be a critical layer separation $d/\ell \sim 2$ below which the QHE is observed.  More recent experiments have explored the transition in samples with extremely weak tunneling ($\Delta_{SAS} \sim 10^{-7}e^2/\epsilon\ell$) and find the critical point to be approximately $d/\ell \approx 1.8$, although there is some variability in this \cite{wellwidth}.  The existing evidence suggests that the energy gap associated with the \nt\ QHE grows smoothly from zero as \dl\ is reduced below its critical value \cite{sawada98,tutuc03,wiersma04}.  Furthermore, the transition between the non-QHE and QHE phases at \nt\ illustrated in Fig. \ref{phasediag} appears to be coincident with the onset of the more exotic transport phenomena (giant zero bias tunneling conductance \cite{spielman00}, vanishing counterflow Hall resistance \cite{kellogg04,tutuc04,wiersma04}, etc.) which are the main subjects of this article.

The precise nature of the transition between the non-QHE phase of the \nt\ bilayer system at large \dl\ and the excitonic QHE phase at small \dl\ remains poorly understood.  Numerous experiments have examined the phase diagram and how it depends on tunneling strength, temperature, layer density imbalance, and spin Zeeman energy and other parameters \cite{murphy94,kellogg03,champagne08a,sawada98,tutuc03,spielman04,clarke05,champagne08b,spielman05,kumada05,giudici08,finck10,luin05,karmakar07,karmakar09,faniel05}. While much has been learned from these experiments, they unfortunately lie outside our present scope.

\section{Condensate Dynamics}

\subsection{Josephson-like Tunneling}
Interlayer tunneling in the coherent \nt\ bilayer has by now been studied extensively, both in experiment \cite{spielman00,spielman01,eisenstein03,spielman04,spielman05,champagne08a,misra08,finck08,champagne08b,tiemann08a,tiemann09,yoon10,nandi12,huang12,nandi13} and theory \cite{wen93,ezawa93,wen96,balents01,stern01,fogler01,joglekar01,iwazaki03,fertig03,bezuglyj04,abolfath04,wang04,jack04,jack05,klironomos05,wang05,bezuglyj05,fertig05,rossi05,khomeriki06,park06,fil07,eastham09,su10,sun10,eastham10,lee11,hyart11,ezawa12}.  The strong tunneling signature has proven to be an effective tool for studying the coherent excitonic phase at small \dl\ and the transition from it to the incoherent compressible phase at larger effective layer separation.

\begin{figure}
\begin{center}
\includegraphics[width=4in] {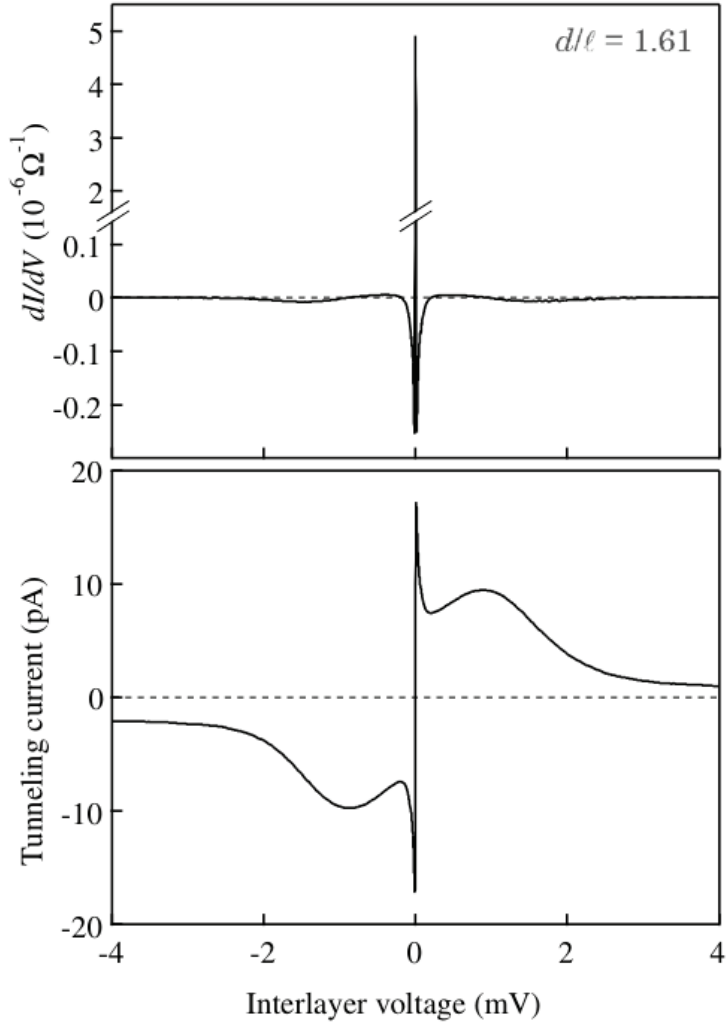}
\end{center}
\caption{Interlayer tunneling at \nt. Upper panel shows the tunneling conductance $dI/dV$ vs. interlayer bias voltage $V$ while the lower panel shows the tunneling current $I$ vs. $V$.  These data were obtained at $T=25$ mK and $d/\ell = 1.61$.  These data reveal a near discontinuity in the tunnel current around $V=0$ and a maximum, or critical, tunneling current of roughly $I_c = \pm 17$ pA.  Immediately beyond this ``supercurrent'' branch the tunneling current drops sharply.  The broad extrema in the current near $V = \pm 1$ mV are due to incoherent tunneling processes.  After Spielman {\it et al.} \cite{spielman01}}
\label{tunnel2}
\end{figure}

Figure \ref{tunnel2} shows interlayer tunneling data at \nt\ taken at low temperature ($T = 25$ mK) and small effective layer separation (\dl=1.61); these conditions place the system well within the coherent excitonic phase.  The differential tunneling conductance $dI/dV$ shown in the top panel reveals an extremely sharp peak centered at zero bias.  The height of the peak falls steadily with increasing temperature and layer separation \cite{spielman00,spielman01,eisenstein03}. (In fact, these dependences define a phase boundary in the $T-d/\ell$ plane \cite{champagne08a}.)  The width of the peak (about 6 $\mu$V for the data in Fig. \ref{tunnel2}) grows steadily with increasing temperature.

The current-voltage ($IV$) data shown in the lower panel of Fig. \ref{tunnel2} are suggestive of the dc Josephson effect \cite{josephson62}.  The device appears to allow tunneling currents up to maximum, or critical value (about 17 pA here) with virtually no voltage appearing across the junction. Beyond this ``supercurrent'' branch the tunneling current falls and significant interlayer voltage develops.  This resistive portion of the $IV$ characteristic exhibits two distinct regions.  First, there is the very rapid fall of the tunneling current just beyond zero bias.  This region corresponds to the deep negative differential conductance spikes in the $dI/dV$ data shown in the upper panel, and is analogous to the resistive state of a superconducting Josephson junction with a time-varying phase difference $\Delta \phi$ across it.  Second, there is the broad peak in the tunneling current at relatively high voltages ($\sim 1$ mV).  This second feature in the $IV$ reflects high-energy incoherent tunneling processes which have little to do with the coherent excitonic ground state of the bilayer.  This broad peak in the $IV$ remains virtually unchanged as \dl\ is increased beyond the critical point and the bilayer becomes essentially two independent 2D electron systems \cite{eisenstein03}.

The data shown in Fig. \ref{tunnel2} were obtained from a simple two-terminal conductance measurement.  As such, they include the effects of all resistances in series with the tunnel junction itself.  At the most fundamental level there are ``contact resistances'' of order $h/e^2$ associated with the injection of current into the bilayer quantum Hall state \cite{pesin11}.  These quantum Hall contact resistances cannot be removed, and in fact eliminate the possibility of observing a true dc Josephson effect in the {\it two-terminal} $IV$ characteristic.

More mundane effects also contribute to the net series resistance $R_{series}$.  In particular, significant resistances can be encountered in the 2D electron gas ``arms'' which lead from the ohmic contacts to the gated portion of the device where the coherent excitonic phase exists.  In the end, $R_{series}$ can be large; $\sim 100$ k$\Omega$ is not uncommon.  This series resistance has various consequences.  First, it obviously limits the maximum observable two-terminal tunneling conductance to $R_{series}^{-1}$.  Second, the series resistance exaggerates the apparent width of the tunneling resonance \cite{width}.  Both of these effects are present to some degree in the data shown in Fig. \ref{tunnel2}.

The series resistance can also create instabilities and hysteresis in the tunneling $IV$ curve.  These instabilities first appear when the negative differential resistance of the junction precisely cancels the positive series resistance in the external circuit \cite{instab}.  This is a common phenomenon among circuits (e.g. tunnel diode oscillators) containing strongly non-linear elements.  In electrical engineering language, one says that the instabilities occur when the ``load line'' (determined by the voltage source and series resistances) intersects the intrinsic $IV$ curve of the non-linear element at more than one point.  For the data in Fig. \ref{tunnel2} this does not occur since the $least$ negative differential resistance is $\sim -4$ M$\Omega$, while the series resistance is only about 100 k$\Omega$.  As a result, the $IV$ curve remains single-valued.  However, recent experiments \cite{tiemann08a,tiemann09,nandi12,huang12,nandi13} on devices with much larger tunneling conductances clearly display the instabilities.

\begin{figure}
\begin{center}
\includegraphics[width=6in] {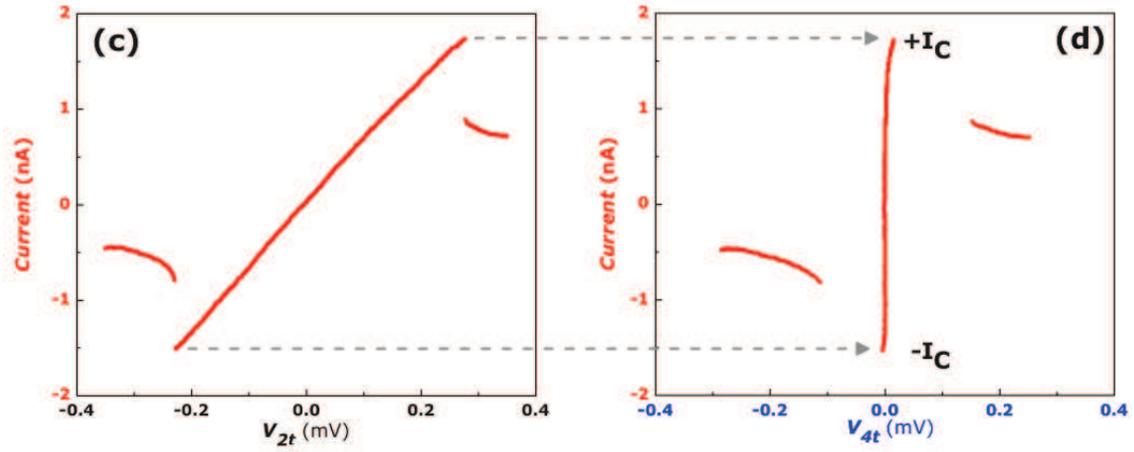}
\end{center}
\caption{Tunneling at \nt. Left panel:  Two-terminal $IV$ characteristic.  Linear segment around $V=0$ implies a series resistance of roughly 150 k$\Omega$.  Instability at $\pm 1.6$ nA is clearly evident. Right panel: Four-terminal $IV$ reveals the Josephson-like jump in the tunneling current at $V_{4 pt} = 0$.  After Tiemann {\it et al.} \cite{tiemann09}}
\label{2v4}
\end{figure}

In order to avoid the distorting effects of the series resistance and more faithfully expose the intrinsic $IV$ of the tunnel junction, a four-terminal technique must be employed.  Two additional ohmic contacts, one on each 2D layer, are used to directly sense the interlayer voltage present when tunneling current flows.  Figure \ref{2v4} clearly displays the difference between the two- and four-terminal $IV$ characteristics of a strongly tunneling \nt\ device.  The figure shows both the distortion of the two-terminal $IV$ curve and the instability near the critical current.  Owing to the instability, the four-terminal voltage jumps discontinuously between essentially zero and a finite value as the critical current is approached.  As a result, important region of the intrinsic four-terminal $IV$ is excluded from measurement in strongly tunneling devices \cite{hyart11,nandi13}.  

The data shown in Figs. \ref{tunnel2} and \ref{2v4} suggest the existence of a maximal, or critical current $I_c$ for interlayer tunneling in the coherent \nt\ bilayer system.  Experiments \cite{eisenstein03,tiemann08a,tiemann09,nandi13} have shown that $I_c$ grows continuously from zero as \dl\ is decreased below a critical value of about $(d/\ell)_c \sim 1.8$.  At elevated temperatures the slope of the four-terminal $IV$ curve at zero bias becomes less steep and the maximum tunneling current occurs at finite voltage.  This maximum current falls steadily as the temperature is increased.  

\begin{figure}
\begin{center}
\includegraphics[width=5in] {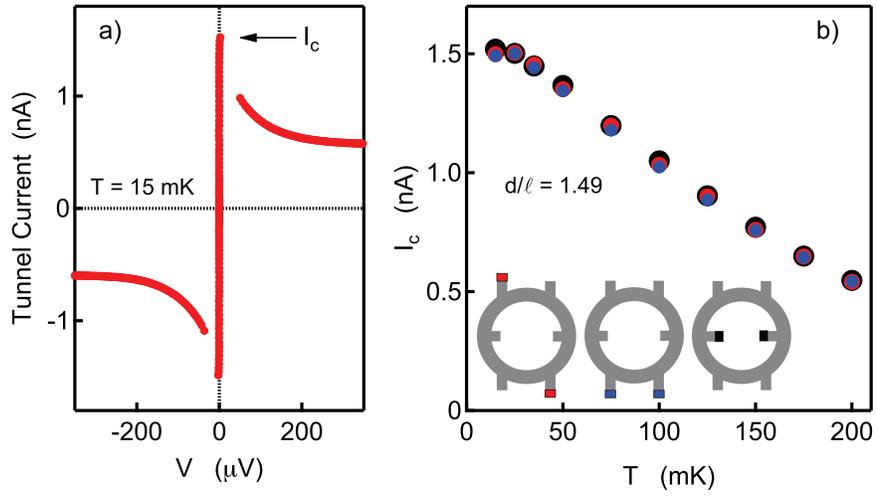}
\end{center}
\caption{a) Four-terminal tunneling $IV$ curve at \nt\ from a large area Corbino device, with critical current $I_c$ indicated.  For these data \dl\ = 1.49 and $T = 15$ mK.  b) Temperature dependence of critical current for three different contact configurations on the same device.  Insets schematically depict arrangements, with colored squares indicating source and drain contacts (with one on the top 2D layer and one on the bottom 2D layer).  The various configurations yield the same critical currents. After Nandi {\it et al.} \cite{nandi13}}.
\label{tunnel3}
\end{figure}

There is by now substantial experimental evidence that interlayer tunneling in the \nt\ bilayer QHE state is uniformly spread across the area of the device \cite{finck08,tiemann09,huang12,nandi13,areadep}.  For example, recent experiments using Corbino annular geometries have shown the critical current to be virtually identical independent of how far apart the source and drain contacts are.  Figure \ref{tunnel3} illustrates this with plots of  $I_c$ vs. temperature for three different contact pairs, the distance between which varies by a factor of 3.  While this rules out a scenario in which tunneling is proportional to the perimeter of the \nt\ quantum Hall droplet, it might seem to leave open the possibility that the tunneling is confined to small ``hot spots'' near the source and drain.  This however conflicts with the clearly observed dependence of $I_c$ on the area of the \nt\ droplet and with the very recent remarkable observation that when current is injected and withdrawn from $two$ remote contact pairs, the same critical current applies to the $total$ current injected into the device \cite{huang12,nandi13}.  

The area scaling of the tunneling conductance seems at first surprising since the in-plane conductivity $\sigma_{xx}$ of a 2D electron system is heavily suppressed when a quantum Hall state is present.  Without any in-plane conductivity, tunneling would presumably be confined to the boundaries of the 2D system where the source and drain contacts reside.  In the \nt\ excitonic QHE state this obstacle might be overcome by the unusual condensate transport mechanism it possesses. This mechanism, of course, is exciton, or counterflow transport, and it seems tailor-made to accommodate bulk interlayer tunneling which necessarily involves counter-propagating currents in the two 2D layers.  Unfortunately, this argument, however appealing, is too glib.  In steady state, charge conservation requires that the divergence of the counterflow current density, $\nabla \cdot j_{ex} \sim \rho_s \nabla^2 \phi$, match the tunneling current density $j_t \sim \Delta_{SAS}$ sin$\phi$ (where $\rho_s$ is the pseudo-spin stiffness, \dsas\ the single-particle tunnel splitting, and $\phi$ the condensate phase).  A natural length scale, analogous to the Josephson penetration length, emerges: $\lambda_J = 2 \ell \sqrt{\pi \rho_s/\Delta_{SAS}}$.  As a result, interlayer tunneling currents are confined to within $\lambda_J$ of sample boundaries. Since estimates of $\lambda_J$ are typically in the $\mu$m range, and typical tunneling devices are much larger, this picture conflicts with the experimental evidence for bulk interlayer tunneling.  This is perhaps not surprising since the same line of argument predicts tunneling critical currents which scale linearly with \dsas\ and are far larger than those observed in experiment \cite{Icest}.

The small magnitude of the observed critical currents and their scaling with the area of macroscopic tunnel junctions are two examples of the substantial quantitative discrepancy between the theory of an ideal, disorder-free bilayer \nt\ system and the actual samples available to experimentalists.  An additional notable example is the poorly understood effect of an in-plane magnetic field \bpar\ on the tunnel resonance.  While early tunneling experiments \cite{spielman01}  successfully employed in-plane magnetic fields to identify the expected linearly dispersing Goldstone collective mode arising from the spontaneously broken U(1) symmetry of the pseudo-ferromagnetic, or excitonic, ground state \cite{fertig89,wen92}, the signatures of these modes were very weak.  Instead of the expected splitting of the tunneling conductance peak \cite{balents01,stern01,fogler01}, the dominant effect of the in-plane field was to suppress the peak height while the collective modes appeared only as subtle satellite features.

There have been numerous theoretical efforts to include the effects of disorder on the \nt\ tunneling problem \cite{balents01,stern01,fertig03,abolfath04,jack04,jack05,fertig05,eastham09,sun10,eastham10,hyart11}.  Prominent among the possible types of disorder are statistical fluctuations in the density of the silicon dopants which donate electrons to the double quantum well.  As a result, the density of the 2DES typically varies by several percent \cite{pikus93} on a length scale determined by the distance between the donors and the quantum wells ($d_s \sim 200$ nm).   It is generally believed that due to these imperfections, vortices in the phase field $\phi$ (which carry electrical charge) are present at all temperatures.  The effects of both static and dynamic vortex fields have been considered. The vortices strongly suppress the tunneling conductance and at least qualitatively explain its area scaling and surprising \bpar\ dependence.  It is remarkable that disorder is at once responsible for both the short coherence lengths ($\xi \sim d_s$) implied by the \bpar\ experiments \cite{bpar} and the very long lengths ($\langle \lambda_J \rangle \sim 1$ mm, with $\langle .. \rangle$ indicating disorder averaging) suggested by the area scaling and other ``global'' properties of the tunneling critical current. 

At the lowest temperatures and effective layer separations, the peak four-terminal conductance in the ``supercurrent'' branch of the $IV$, while still apparently finite, has been observed \cite{nandi13} to exceed $\sim 250~e^2/h$.  The corresponding voltage width of the supercurrent branch is then less than 0.3 $\mu$V, considerably less than $k_BT$ at the 30 mK measurement temperature. 
It is worth pausing to note that this zero bias tunneling conductance at \nt\ is more than 5000 times larger than the tunneling conductance observed at zero magnetic field in the same sample.  This comparison vividly contrasts the coherent, many-particle aspect of tunneling at \nt\ with its essentially single-particle character at zero magnetic field \cite{zerofieldtun}.

The ultimate fate of the four-terminal tunneling $IV$ curve in the limit of extremely low temperatures and voltages remains 
unknown.  Is there a true supercurrent branch at $V=0$ on which the phase field $\phi$ is time-independent, or is there always a residual finite tunneling resistance?  Even if this resistance remains non-zero, it is not obvious whether it is intrinsic to the tunneling process or originates instead from the in-plane transport which must be present in order to allow tunneling to occur throughout the bulk of the 2D system \cite{vertical}. 

\subsection{Quantized Hall Drag}
The first experimental indication for unusual in-plane transport properties of coherent \nt\ bilayers came from Coulomb drag experiments performed in simply-connected square geometry \cite{kellogg02,simply}.  In such measurements current is driven through one of the layers while voltage differences are measured in the other, non-current carrying, layer.  At zero magnetic field the drag resistance (the ratio of drag voltage to drive current) reflects momentum transfer due to interlayer electron-electron interactions \cite{gramila91,rojo99}.  Application of a perpendicular magnetic field modifies the longitudinal drag resistance $R_{xx,D}$, but rarely creates a large transverse, or Hall drag resistance $R_{xy,D}$ in weakly coupled bilayers \cite{nohalldrag}.  In general, large Hall drag resistances are only expected when strong interlayer electronic correlations exist \cite{renn92,moon95,duan95,yang98,yang01,kim01}.  

\begin{figure}
\begin{center}
\includegraphics[width=4in]{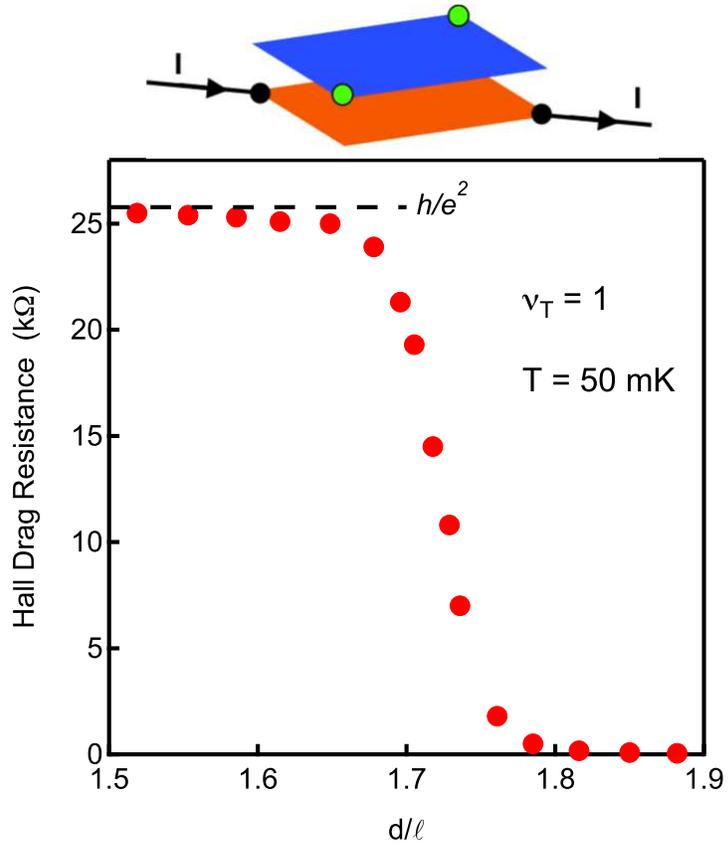}
\end{center}
\caption{Quantization of Hall drag at \nt. Drawing depicts idealized Hall drag set-up in a simply-connected, square geometry.  Current flows only through lower layer, while Hall voltage is measured across opposite layer (via the green contacts).  Data shows development of Hall drag resistance quantization to $R_{xy,D} = h/e^2$ as \dl\ is reduced.  After Kellogg {\it et al.} \cite{kellogg02,kellogg03}.}
\label{halldrag}
\end{figure}

Experiments on \nt\ bilayers reveal that at low temperatures the Hall drag resistance smoothly rises from essentially zero at large \dl\ in the weakly coupled compressible phase to become accurately quantized at $R_{xy,D} = h/e^2$ in the incompressible excitonic phase at small effective layer separations \cite{kellogg02,kellogg03,tutuc04,wiersma04,tutuc09}.  This behavior is shown in Fig. \ref{halldrag}. It is remarkable that in the excitonic phase the same Hall voltage is observed in both layers at \nt, even though only one of them is carrying a net current. 

The existence of equal quantized Hall voltages in the two layers suggests that equal quasiparticle currents are flowing in the two layers in spite of the drag circuit setup which is designed to drive current only through one of the layers.  While current leakage due to strong interlayer tunneling might at first seem sufficient to defeat the drag boundary condition and thereby explain the Hall drag result, this has been convincingly shown not to be the case \cite{halldrag}.  

The quantization of the Hall drag at \nt\ finds a ready explanation within the standard theoretical picture of the excitonic phase.  In order to satisfy the drag boundary condition and yet simultaneously produce a quantized Hall voltage across the drag layer, a neutral exciton current in the condensate must accompany the charged quasiparticle current which is equally shared between the layers.  The net current $I$ injected into the drive layer is resolved into a symmetric quasiparticle current of magnitude $I$ (half of which flows in in each layer) and a counterflow current of the same magnitude in the exciton condensate.  The counterflow current adds to the quasiparticle current to produce the total current $I$ in the drive layer but subtracts from it to yield zero net current in the drag layer.  Aside from the neutral exciton current, the situation is now no different from an ordinary Hall effect measurement on the two layers in parallel; exact quantization of the Hall voltage appears across $both$ layers.  In this way, the quantization of the Hall drag resistance provides indirect evidence for neutral exciton transport in the \nt\ condensate.

\subsection{Counterflow in Hall Bar Geometry}

\begin{figure}
\begin{center}
\includegraphics[width=4in]{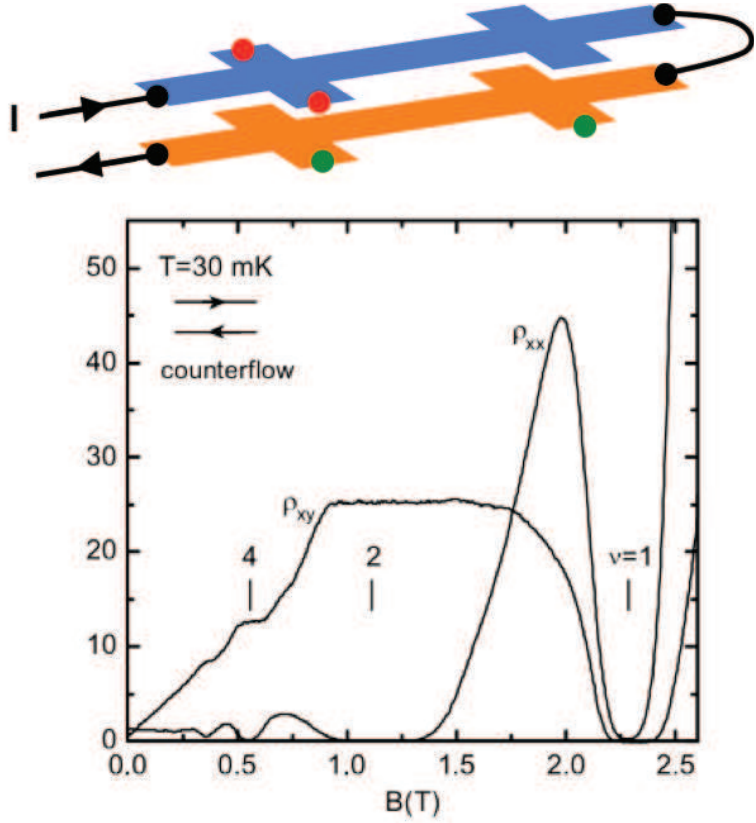}
\end{center}
\caption{Counterflow transport in Hall bars.  Drawing at top shows idealized geometry, with current flowing left to right in top layer and right to left in bottom layer.  Hall voltage is measured between red contacts, longitudinal voltage between green contacts.  Data shown is for a bilayer 2D hole system.  Both Hall and longitudinal resistances collapse around $B = 2.3$ T where \nt.  After Tutuc {\it et al.} \cite{tutuc04}.}  
\label{HBCF}
\end{figure}

A natural extension of the Hall drag measurements is to route the drive current through both 2D layers but in opposite directions.  This constitutes a counterflow setup and therefore seems ideal for generating transport within the neutral exciton condensate without simultaneously coupling to the charged degrees of freedom as Coulomb drag measurements do.  As already shown via Fig. \ref{thumbs}, such Hall bar counterflow measurements yield dramatic results, notably the vanishing of the Hall voltage across each of the two 2D layers at \nt.  Measurements on both bilayer 2D electron and bilayer 2D hole systems at \nt\ reveal that the longitudinal voltage also appears to vanish in the low temperature limit \cite{kellogg04,tutuc04,wiersma04,tutuc05,yoon10}.  Hence, we are confronted by the remarkable fact that all four components of the counterflow resistivity $\rho_{ij}^{CF}$ tensor vanish as $T \rightarrow 0$.  Figure \ref{HBCF} shows early counterflow transport data in a bilayer hole system which illustrate this fact. 

The above results on counterflow transport in Hall bar geometries are qualitatively consistent with the counterflowing currents being transported by excitons in the \nt\ condensate.  However, with contacts only on the outside edge of the device,  Hall bar experiments cannot directly demonstrate that the putative excitonic currents are, as expected, free to  move through the {\it bulk} of the 2D system.  For this, we turn to the very recent measurements in Corbino geometries which settle this question unambiguously.

\subsection{Counterflow in Corbino Geometry}
Quantum Hall systems are topological insulators. They are electrical insulators in the bulk but possess topologically protected conducting edge states at their boundaries.  Transport measurements made using Hall bar geometries, for which all contacts to the 2D system lie along a single outer boundary, reveal that the longitudinal resistivity $\rho_{xx}$ vanishes at low temperatures when a strong QHE state is present.  Such measurements indirectly demonstrate that the bulk charge conductivity $\sigma_{xx}$, deduced by inverting the resistivity tensor, also vanishes.  A more direct demonstration of the vanishing bulk conductivity is obtained using Corbino multiply-connected geometries, a simple example of which is an annulus.  While conducting edge channels are present on each rim of the annulus, there is no conducting pathway between the two rims.  The conductance measured between the two rims, proportional to $\sigma_{xx}$, is exponentially small at low temperatures.  This same result applies to the bilayer \nt\ QHE state, provided the measurement is set up to drive equal currents in the same direction through the two layers.  Tiemann {\it et al.} \cite{tiemann08b} were the first to make Corbino measurements on the \nt\ bilayer system. 

While transport of charged quasiparticle excitations across the bulk of the bilayer \nt\ QHE state is suppressed just as it is in all QHE states, the existence of the condensate degree of freedom in the coherent bilayer allows for an independent form of bulk transport.   Gradients in the condensate phase $\phi$ correspond to neutral currents which may be viewed as equal, but oppositely directed charge currents in the two layers or, equivalently, as exciton currents.  One thus distinguishes, in the \nt\ bilayer, between the parallel current charge conductivity \Sxx, which is extremely small, and the counterflow, or exciton conductivity \SxxCF, which is expected to be extremely large, if not infinite.  

\begin{figure}
\begin{center}
\includegraphics[width=6in] {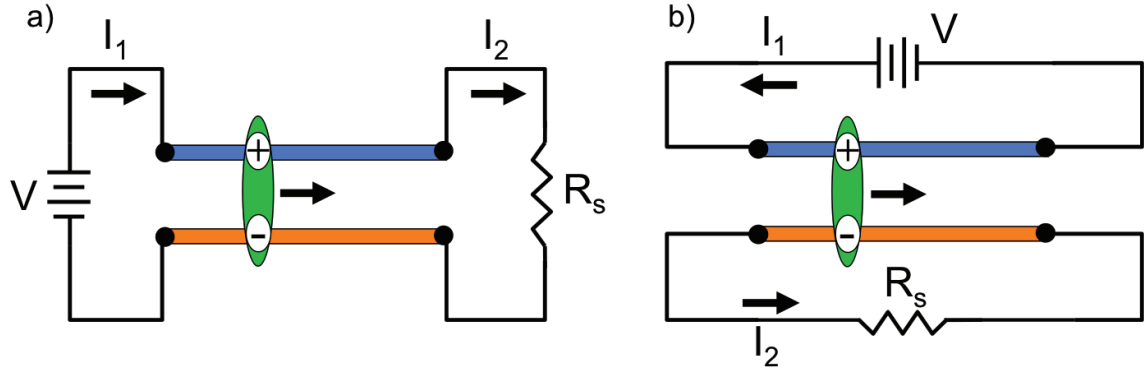}
\end{center}
\caption{Schematic circuits for detecting exciton transport in Corbino geometry.  a) Series counterflow (SCF). b) Drag counterflow (DCF).  Drawings represent cross-sections through the Corbino annulus, with the contacts (black dots) on the left being on one rim of the annulus and those on the right being on the opposite rim.  No conducting edge channels connect the contacts on the left with those on the right.  In both cases, pure exciton transport requires the currents $I_1$ and $I_2$ to be equal. Both sketches omit the inevitable extrinsic series resistances present in actual experimental set-ups.}
\label{circuits}
\end{figure}

As noted by Su and MacDonald \cite{su08}, there are two basic circuits, dubbed ``series counterflow'' (SCF) and ``drag counterflow" (DCF), for generating excitonic flow across the bulk of a Corbino annulus at \nt.  These two geometries are schematically illustrated in Fig. \ref{circuits}. In the SCF case a voltage $V$ is applied between contacts to the two separate layers on one rim of the annulus.  Meanwhile, on the opposite rim the two layers are connected together via a shunt resistor, $R_s$.  Aside from this shunt connection, this is a two-terminal tunneling set-up. Alternatively, in the DCF case, the voltage is applied between contacts on the same layer, but on opposite rims, while the shunt resistor $R_s$ connects the inner and outer rim via contacts on the opposite layer.  In this DCF geometry there is no explicit connection between the layers and the configuration closely resembles that used for measurements of Coulomb drag.  (Keep in mind that there are always substantial series resistances, not shown in Fig. \ref{circuits}, between the coherent \nt\ bilayer 2DES and external circuit elements.)  In the ideal situation, where only exciton transport is important, the current $I_1$ supplied by the battery and the current $I_2$ flowing through the shunt will be precisely equal in both the SCF and DCF circuits.  Furthermore, if the exciton transport is truly dissipationless, the magnitude of this current would be determined entirely by the battery voltage and the sum of all series resistances (including quantum Hall contact resistances of order $h/e^2$) in the circuit.  However, this ideal scenario is affected by interlayer tunneling and non-zero charge conductivity \Sxx\ in important ways that we now briefly discuss.

The strong Josephson-like interlayer tunneling characteristic of the coherent \nt\ bilayer influences transport measurements in both the SCF and DCF circuits.  In the SCF case very little current will flow through the shunt resistor $R_s$ until the battery has driven the tunnel junction into the resistive state.  Prior to that point the total current supplied by the battery is less than the tunneling critical current and there is essentially no voltage drop between the layers or across the shunt resistor.  In this regime there is no net counterflow, or exciton transport, crossing the bulk of the Corbino annulus.  In the DCF case, strong interlayer tunneling enables current to flow through the shunt resistor $R_s$ even if there is no exciton transport across the bulk of the Corbino ring.

The parallel conductivity \Sxx\ is never truly zero in a quantum Hall state.  Both finite temperatures and non-linear effects at finite source-drain voltage render \Sxx\ non-zero, even if quite small.  In the SCF set up shown in Fig. \ref{circuits}a, the circuit itself prevents any net current from flowing between the two rims of the Corbino annulus.  Hence, finite parallel conductivity does not directly influence SCF experiments.  This contrasts with the DCF case where the circuit itself does not prohibit a net current flow between the two rims.  A finite \Sxx\ definitely affects the outcome of DCF experiments, especially at elevated temperatures and drive levels.

In the experiments by Finck {\it et al.} \cite{finck11} the SCF configuration was explored in a Corbino annular geometry.  As expected, no current was observed to flow through the shunt resistor $R_s$ until the battery had driven tunnel junction into its resistive state.  Beyond this point the shunt current began to grow and approach the total current supplied by the battery.  The same basic effect was observed previously by Yoon {\it et al.} in counterflow experiments in a Hall bar geometry \cite{yoon10}.

In order to render bulk exciton transport the dominant transport process, Finck {\it et al.} tilted their Corbino sample by $\theta = 28^{\circ}$ relative to the magnetic field direction.  The field component $B_\perp$ perpendicular to the 2DES was kept the same as in the untilted state, thus maintaining \nt\ and $d/\ell=1.5$.  As expected \cite{spielman01}, the tilt-induced in-plane magnetic field component ($B_{||} \approx 1$ T) heavily suppressed the coherent tunneling resonance, reducing the zero bias critical current from about 1.5 nA down to only $I_c \approx 5$ pA.  With this arrangement, Finck {\it et al.} \cite{finck11} found that the shunt current $I_2$ and the battery current $I_1$ were virtually identical and grew in approximately linear \cite{nonlinear} proportion to the battery voltage $V$. Independent measurements (on the same sample) showed that the ratio of current to voltage in this experiment was due essentially entirely to the various series resistances in the circuit; whatever dissipation might be occurring in the exciton transport itself was not detectable.

The Corbino experiments of Finck {\it et al} \cite {finck11} demonstrated that counterflowing electrical currents could readily cross the insulating bulk of the bilayer 2DES.  This is a remarkable finding, given the fact that parallel layer currents, which transport net charge, encounter enormous resistance (proportional to $1/$\Sxx) under the same conditions. Finck {\it et al.} further showed that even when an external low resistance pathway from the shunt resistor back to the battery was provided, very little current flowed through it, thus demonstrating that the high counterflow conductivity is an intrinsically bilayer effect.

The conclusion that counterflowing, or excitonic, currents could cross the bulk of the \nt\ bilayer 2DES was not possible based on counterflow experiments done in Hall bar geometries \cite{kellogg04,tutuc04,wiersma04,yoon10} in which all ohmic contacts reside on the single outside edge of the 2DES.  Furthermore, uncertainty about the role of the conducting edge channels is effectively removed in the Corbino geometry.

The first drag counterflow (DCF) experiments in Corbino geometry were performed by Tiemann {\it et al.} \cite{tiemann08a}.  With the coherent \nt\ quantum Hall phase well established in the Corbino annulus, the observed drive and drag currents ($I_1$ and $I_2$) were of equal magnitude and, as expected, oppositely directed.  However, Tiemann {\it et al.} noted that it remained unclear to what extent the observed drag current was due to bulk exciton transport across the Corbino annulus as opposed to the Josephson-like interlayer tunneling characteristic of the coherent \nt\ phase \cite{tiemann08a}.  Subsequent DCF experiments by Nandi {\it et al.} \cite{nandi12} employed tilted magnetic fields to suppress the tunneling and allow for a clear demonstration of exciton-mediated Coulomb drag.

\begin{figure}[t]
\begin{center}
\includegraphics[width=4.5in] {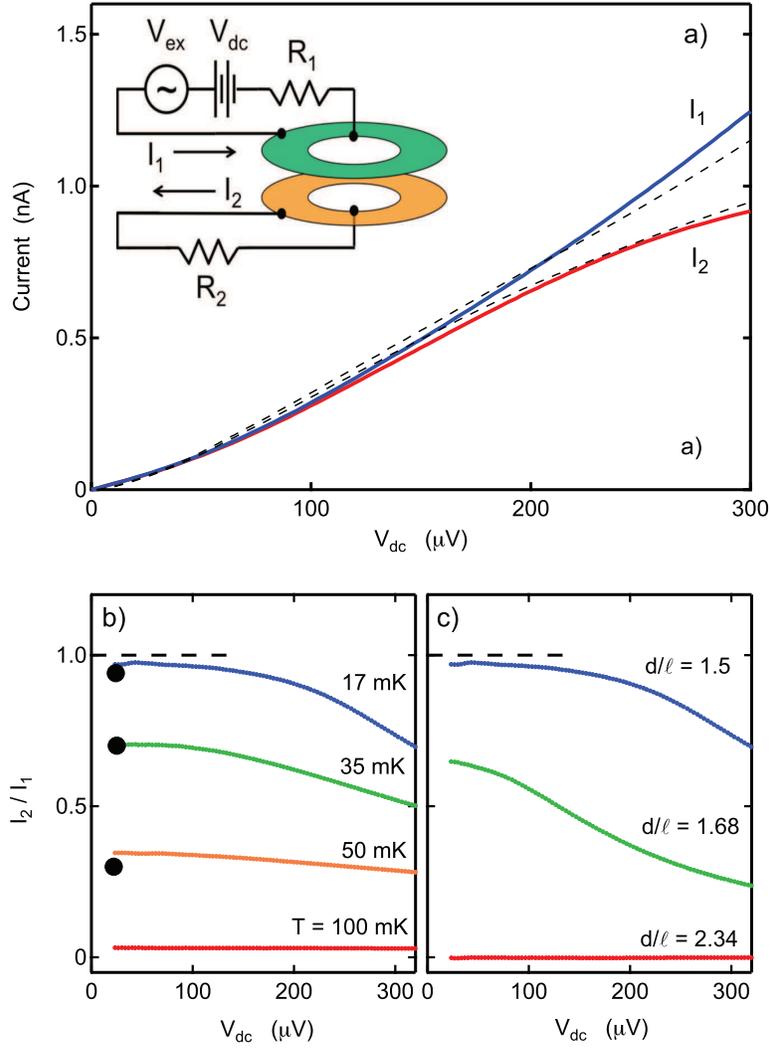}
\end{center}
\caption{Drag counterflow measurements at \nt.  a) $I_1$ and $I_2$ vs. $V_{dc}$ at $T=17$ mK and $d/\ell=1.5$ where the excitonic phase is well established.   The inset schematically illustrates the experimental arrangement.  The resistances $R_1$ and $R_2$ represent the total series resistances in each loop.  For these data the ac excitation voltage $V_{ac}$ was set to zero.  Note that the observed currents $I_1$ and $I_2$ are oppositely directed in the two layers of the Corbino annulus.  b) Ratio of drag to drive current $I_2/I_1$ vs. $V_{dc}$ for several temperatures.  c) $I_2/I_1$ vs. $V_{dc}$ at different $d/\ell$. After Nandi {\it et al.} \cite{nandi12}.}
\label{corbinodrag}
\end{figure}

Figure \ref{corbinodrag}(a) shows the drive and drag currents observed by Nandi {\it et al.} deep within the excitonic \nt\ phase.  For dc drive voltages below about $V_{dc} \sim 150$ $\mu$V the drag is essentially ``perfect'', i.e. $I_1 = I_2$.  As expected, the currents are oppositely directed in the two layers.  Since the observed currents are far larger than the maximum tunneling currents measured under the same conditions, Nandi {\it et al} concluded that their results were dominated by exciton transport across the bulk of the Corbino ring, with interlayer tunneling playing a negligible role.  Unlike the SCF results of Finck {\it et al.} \cite{finck11}, these DCF experiments demonstrate bulk exciton transport without any explicit electrical connection between the two layers. 

As Fig. \ref{corbinodrag}(a) shows, the drive and drag currents $I_1$ and $I_2$ become unequal at elevated $V_{dc}$.  Similarly, raising the temperature or the effective layer separation \dl\ at \nt\ also renders the drag imperfect, even in the limit $V_{dc} \rightarrow 0$.  These effects are demonstrated in Figs. \ref{corbinodrag}(b) and \ref{corbinodrag}(c) where the ratio $I_2/I_1$ is plotted vs. $V_{dc}$ for various temperatures and $d/\ell$.  Obviously, when $I_1 \neq I_2$, net current is being transported across the bulk of the Corbino annulus.  Hence, these departures from perfect Coulomb drag are rooted in non-zero values of the ordinary charge conductivity \Sxx.  Nandi {\it et al.} were able to successfully model this effect by incorporating independent measurements of the temperature, $d/\ell$, and $V_{dc}$ dependences of \Sxx.  The dashed lines in Fig. \ref{corbinodrag}(a) and the solid dots in Fig. \ref{corbinodrag}(b) are the results of such modeling.

\subsection{How Super a Superfluid?}

In the absence of tunneling and disorder, the bilayer \nt\ exciton condensate is expected to be a 2D superfluid, with counterflowing electrical currents able to flow with little or no dissipation.  In this idealized scenario, a finite temperature Kosterlitz-Thouless (KT) phase transition is expected \cite{wen92,yang94,moon95} in the \nt\ system, with vortices in the condensate phase $\phi$ binding up into pairs at temperatures below a critical temperature $T_{KT} \sim \pi \rho_s/2$.  While free vortices create dissipation in the presence of uniform exciton transport (i.e. counterflow), bound vortex-antivortex pairs do not.  Counterflow transport will exert Magnus forces on the vortices and lead to ionization of the pairs, and thus dissipation.  The effective current-voltage characteristic for counterflow transport is expected to be highly non-linear: $V \sim I_{CF}^p$, with the exponent $p$ jumping from $p=1$ to $p=3$ as the temperature falls below $T_{KT}$, and rising steadily as the temperature is reduced further \cite{moon95}.  Hence, truly dissipationless transport is only expected in the $I_{CF} \rightarrow 0$ limit.

Unlike the situation in other 2D superfluids (notably thin helium films \cite{bishop78,agnolet89}), clear-cut signatures of KT physics in the \nt\ bilayer system have remained elusive.  Evidence for a critical temperature has been reported, but whether it reflects a genuine KT transition is uncertain \cite{lay94,champagne08c,terasawa12}.  While it is clear from experiment that dissipation in counterflow transport is quite small, the Hall bar data suggests the presence of a small, linear resistivity $\rho_{xx}^{CF}$ at low temperatures and effective layer separations.  Indeed, the Hall bar measurements show that the temperature dependence of $\rho_{xx}^{CF}$ is not unlike the ordinary resistivity $\rho_{xx}$ of any ``dissipationless'' quantized Hall state, with values as low as $\rho_{xx}^{CF} \lesssim 50$ $\Omega$ having been reported \cite{kellogg04,tutuc04,wiersma04}. (Corbino counterflow measurements have yet to reach a comparable sensitivity to dissipation.)

It seems likely that this residual linear dissipation is due to the existence of unpaired vortices at low temperatures \cite{fertig03,jpe04,huse05,fertig05,eastham09}.  Vortices in the \nt\ system carry electrical charge ($\pm e/2$) and may be nucleated by the disorder potential arising from statistical fluctuations in the dopant population.  The motion of such vortices could explain the observed linear dissipation in counterflow.  The non-linear KT vortex pair ionization mechanism presumably also exists, and improvements in sample quality and measurement techniques might yet reveal it. 

\section{Conclusions and Outlook}
It is abundantly clear that bilayer 2D electron systems at total filling factor \nt\ condense into an unusual state of quantum electronic matter when the separation between the layers is sufficiently small and the temperature is sufficiently low.  The electron system is then incompressible, exhibiting a quantized Hall effect even when the tunneling rate between the layers is arbitrarily small.  Moreover, transport experiments (e.g. interlayer tunneling, counterflow, etc.) which are antisymmetric in the layer degree of freedom yield especially dramatic results.  These results clearly expose the presence of an underlying exciton condensate which is capable of nearly dissipationless transport throughout the bulk of the system.  This transparency to neutral exciton transport contrasts sharply with the system's robust opacity to bulk charge transport.   

Although the overall theoretical understanding of the bilayer \nt\ system is well advanced, important unanswered questions remain.    Though not addressed here, the precise nature of the transition into the excitonic phase is one of these questions. Perhaps most importantly, the way in which disorder affects the transport properties of the exciton condensate is not well understood.  The consequences of disorder are far from subtle: Tunneling critical currents are orders of magnitude smaller than expected and yet exhibit unexpected (and intriguing) global properties.  Dissipation in counterflow (i.e. exciton) transport is small, but apparently non-zero and linear. No evidence for the expected non-linear Kosterlitz-Thouless effects has yet been found.  

There are numerous avenues for future experimental work.  For example, analogs to the $ac$ Josephson effect in superconducting junctions should exist in the bilayer \nt\ system \cite{wen93,ezawa93,stern01,hyart13}.  In-plane tunnel junctions between coherent \nt\ droplets are predicted to exhibit an excitonic version of the Josephson effect \cite{wen96,park06}.  More precise studies of counterflow transport may yet reveal the expected Kosterlitz-Thouless physics.  Will the \nt\ exciton condensate lead to thermal transport anomalies analogous to those found in superfluid helium?  Why has an excitonic state not yet been found at $\nu_T =3$ and $\nu_T = 1/3$?  And what about exciton condensation in other physical systems (e.g. double layer graphene, thin film topological insulators, etc.)?  Some of these questions will eventually be answered.  No doubt beguiling new puzzles will pop up along the way.

\break
{\bf Acknowledgements}
My thanks go to my students and postdocs Alex Champagne, Aaron Finck, Mindy Kellogg, Trupti Khaire, Debaleena Nandi, Ian Spielman, and Lisa Tracy for performing the Caltech portion of the experiments described here. They and I are indebted to Loren Pfeiffer and Ken West for growing numerous high quality GaAs/AlGaAs heterostructures in support of our work. I also wish to thank Sankar Das Sarma, Steve Girvin, Gil Refael, Ady Stern, and Kun Yang for innumerable very helpful discussions.  Allan MacDonald deserves special thanks for his sustained interest in the physics described here and his infinite patience in explaining so much of it to me.  The Caltech work has been generously supported by grants from the National Science Foundation, the Department of Energy, and the Gordon and Betty Moore Foundation.  Finally, it is a pleasure to thank the Tata Institute of Fundamental Research in Mumbai and the Indian Institute of Science in Bangalore for their kind hospitality while this article was being written.

\clearpage

\clearpage

\clearpage

\clearpage

\clearpage

\clearpage

\clearpage

\clearpage

\clearpage

\clearpage


\begin{references}
\bibitem{pstheory} There is by now an enormous theoretical literature on quantum Hall ferromagnetism and on its application to the \nt\ bilayer problem.  Refs. \cite{fertig89,wen92,yang94,moon95,smg_ahm97} provide an excellent starting point.
\bibitem{spinwavegap} In fact, for non-zero tunneling the dispersion is gapped at $q=0$.  This gap has so far been too small to be resolved in experiments.
\bibitem{exact} Assuming Landau level mixing can be completely ignored.  Obviously, if the layers are so close together as to be indistinguishable, then Eq. \ref{eq1} is no different than that which describes a $single~layer$ 2D electron system at $\nu =1$.
\bibitem{spin1} At very large tunneling, the \nt\ QHE will be dominated by the spin splitting of the lowest Landau level.   Exchange effects will enhance both the tunneling and spin-flip gaps and influence the cross-over point.
\bibitem{wellwidth} The critical effective layer separation is set by a competition between interlayer and intralayer Coulomb interactions.  As such, it is affected by the thickness of the quantum wells as well as their center-to-center separation.  For the experiments described in this article, the quantum well width is generally between 15 and 20 nm.  See also Ref. \cite{areadep}.
\bibitem{width} For example, a two-terminal measurement of an ideal Josephson junction would display a finite voltage width to the supercurrent branch equal to $R_{series}I_c$, where $I_c$ is the junction critical current.
\bibitem{vertical} Dissipation in the tunneling process alone might be detectable if current could be fed into the tunnel junction $vertically$, say via additional heavily-doped conducting layers above and below the bilayer itself.  This scenario was envisioned in the theoretical work of Park and Das Sarma \cite{park06}.
\bibitem{simply} We here do not distinguish between genuine Hall bars and any other simply-connected geometry with contacts only on the outer boundary.
\bibitem{nohalldrag}  The simple, though incomplete, argument for this is simply that with no current flowing in the drag layer there is no net Lorentz force on the carriers.
\bibitem{halldrag} The quantization of the Hall drag resistance remains unchanged even when the Josephson-like interlayer tunneling is quenched by the application of large in-plane magnetic fields or dc interlayer voltages.  See Refs. \cite{kellogg02,finck10}.
\bibitem{nonlinear} In fact, the currents showed some non-linearity at small $V$.  This was found to be a result of non-linearity in the series resistances leading into the Corbino ring, not exciton transport across the bulk of the annulus.
\end{references}
\end{document}